\documentclass[twocolumn,showpacs,preprintnumbers,amsmath,amssymb]{revtex4} 
\usepackage{graphicx}
\usepackage{dcolumn}
\usepackage{bm}

\def\ni{\noindent}
\def\sig{\sigma}
\def\br{\bf{r}}

\def\bff{\bf{f}}

\def\bv{\bf{v}}
\def\bn{\bf{n}}

\def\bF{\bf{F}}

\def\bV{\bf{V}}
\def\bff{\bf{f}}
\def\bFF{\bf{F}}

\def\bnabla{\bf{\nabla}}

\def\tilT{{\tilde T}}

\def\beq{\begin{equation}} 
\def\eeq{\end{equation}}

\begin{document} 


\title{Plug flow formation and growth in da Vinci Fluids}

\author{Moshe Schwartz}

\affiliation{Beverly and Sackler School of Physics and Astronomy, Tel Aviv University, Ramat Aviv 69934, Israel}

\author{Raphael Blumenfeld}
\email[]{rbb11@cam.ac.uk}
\altaffiliation{Also at: Cavendish Laboratory, JJ Thomson Avenue, Cambridge CB3 0HE, UK}
\affiliation{Earth Science and Engineering, Imperial College, London SW7 2AZ, UK}

\date{\today}

\begin{abstract}

A new, da Vinci, fluid is described as a model for flow of dense granular matter. We postulate local properties of the fluid, which are generically different from ordinary fluids in that energy is dissipated by solid friction. We present the equation of flow of such a fluid and show that it gives rise to formation and growth of plug flow regions, which is characteristic of flow of granular matter. Simple explicit examples are presented to illustrate the evolution of plug flow regions. 

\end{abstract}



\maketitle


\ni On length scales much larger than grain size, dry sand appears to flow similarly to ordinary fluids. Apriori, it should be possible to construct continuum flow equations for dense non-cohesive granular materials and indeed such an approach has a long history. Yet, there is currently no agreement on any one set of such equations as a clear favourite \cite{Ra00}. One of the main reasons for the complex behavior of granular materials is that the particles are inherently inelastic and energy dissipation dominates local dynamics. This feature makes conventional hydrodynamics ineffective and instabilities, such as inelastic collapse \cite{ZhKa96}, play a major role. As a result, large assemblies of macroscopic inelastic particles display a combination of both solid and fluid properties \cite{JaNaBe96}. The kinetic theory of so-called dense gases, which takes into considerations dissipation during collision between particles, is only useful for low-concentration granular systems in very rapid flow \cite{Ha83, LuSaJeCh84, AzChMo99, Go99}. For dense flows, when many particles rub against one another simultaneously, not only that formalism breaks down but also the concept of collision is not useful.  Attempts have been made to extend the kinetic theory, introducing empirical stress tensors that take friction into account \cite{Sa83, JoNoJa90, NoJa92, AnJa92}, but the usefulness of this ad-hoc approach is not yet clear in the presence of multi-particle contacts. Much work is based on proposed phenomenological flow rules, relating strain rates and shear stresses \cite{Sa98, LoBo0002}, but there is no agreement on any one such relation.
Suggestions incorporating the idea of non-uniform stress propagation through force chains have also been put forward \cite{Mi9900}, but the theory behind these stress fields is not fully developed yet \cite{Bl04, GeKrBl08, BoClLeOt01}.

\ni A generic feature of granular flow is the formation and growth of plug regions, wherein the material moves as a rigid body \cite{Bi22, Jo70}. Yet, most theoretical studies appear to assume existence of plug regions, rather than analyse the formation and evolution of such regions from flow equations.

\ni The first purpose of this paper is to construct a minimal-model of flow for a dense system of non-cohesive grains from basic physical considerations, keeping to a simple mathematical description. Our second goal is to study quantitatively the formation and growth of plug regions directly from the equation. 

\ni In this paper, we first derive an equation describing the flow of dense granular matter in regions of space, where the velocity field is not uniform. This involves the construction of a new stress tensor that captures the correct physics of particles interacting via hard core repulsion and solid friction. 
We then extend the model to include a description of the formation of plug regions and the dynamics of their expansion. The examples we present illustrate how local minima or maxima in an initial velocity profile play the role of seeds around which plug zones form and expand. Our analysis shows that, at time $t$ shortly after formation, the linear size of a plug zone is proportional to $t^{1/3}$.

\ni To construct our flow equation, let us consider first the difference between ordinary and granular fluids. Conceptually, the repulsive forces between grains are equivalent to interactions between molecules in ordinary fluids. The main difference is that in granular systems particles exert on each other non-central forces that cannot be viewed as stemming from two-body potentials and are due to solid friction \cite{BoTa50}. Solid friction, described first by da Vinci \cite{dV}, Amontons \cite{Am1699} and Coulomb \cite{Co1779}, is an important energy-dissipating mechanism that transforms mechanical energy into heat stored in intra-granular degrees of freedom. This is different from the physics of ordinary viscous fluids. For example, the conventional dissipative term in the Navier-Stokes equation, which also represents conversion of mechanical energy to heat, only feeds energy from macroscopic fluid disturbances to fluctuations on much smaller scales. Indeed, the viscosity term in the Navier-Stokes equation can be viewed as a result of drag forces on individual particles due to collisions with other particles, leading to viscous dissipation that is linear in the spatial derivatives of the velocity. The total force density, which gives the force on a volume element of the fluid due to other elements in its vicinity, is the sum of two terms: a pressure gradient $-\nabla P$ and a dissipative term. In the following, we assume a simple dense granular fluid, which we call da Vinci fluid (dVF), after Leonardo da Vinci, who conducted the first recorded experiments on solid friction.
The main difference between this and conventional fluids is that the dissipation is assumed to result only from internal solid friction, in contrast to the conventional drag forces that depend on relative velocity. The forces on a volume element in a granular fluid is the sum of the direct contact forces (normal and frictional) applied to its grains by grains in neighbor elements. We assume that the force density can be written as

\begin{equation}
\bff =  - \nabla\cdot\sig^{(n)} + F_f  \  ,
\label{eq:Ai}
\end{equation}
where the first term on the right hand side is due to normal intergranular forces applied on a volume element and the second is due to intergranular solid friction. It should be noted that the boundary between volume elements (which contain a very large number of grains) is a conceptual construct, and that the net normal intergranular forces between grains across a boundary is not necessarily  normal to the boundary. Similarly, the net intergranular friction forces between grains across a boundary does not necessarily align tangentially to the boundary.

\ni The continuum solid friction term in a dVF can be derived in several ways. We have worked it out from detailed microscopic considerations, taking normal and solid friction intergranular forces into consideration. This derivation, however, is too long to reproduce here and it will be reported in detail elsewhere. Instead, we give in the following an alternative derivation from general arguments. 
Consider a region in the fluid, where the (non-symmetric) strain rate tensor $\tilT_{ij}=\partial_i v_j$ is non-zero (here and in the following, $\partial_i$ and  $\partial_t$ stand for  $\partial/\partial x_i$ and  $\partial/\partial t$, respectively). 
When the strain rate does not vanish, the friction force between adjacent volume elements should be proportional to the normal stress tensor, $\sig^{(n)}$. The friction force should also be independent of the relative velocity  between the elements \cite{Co1779}.
The simplest solid friction term consistent with this picture is $-\gamma\nabla\cdot S$, where $S$ is the symmetric part of $[\sig^{(n)}\cdot T]$, $\gamma$ is proportional to an effective inter-granular solid friction coefficient $\mu$, and $T$ is a unit tensor, defined by $T_{ij}=\tilT_{ij}/\mid \tilT\mid$, where $\mid \tilT\mid$ is the norm of $\tilT_{ij}$.

\ni Let us consider first the equation of motion when the flow is nowhere spatially uniform. To make progress we write the 'normal' stress tensor in a simple form

\begin{equation}
\sig^{(n)}_{ij}(\br) = \sig^0_{ij} + P(\rho(\br)) \delta_{ij} \ ,
\label{eq:Aii}
\end{equation}
where $P$ is a scalar, $\rho(\br)$ is the local density and $\sig^0_{ij}$ is position-independent, representing external and body forces. Since the material cannot be compressed to a point, the pressure has to diverge at some maximal density $\rho_c$. For the dense flows we aim to model, the average density, $\bar{\rho}$, is not much lower than $\rho_c$,  $\delta\rho=\rho_c-\bar{\rho}<<\rho_c$. 
A full description of the system is obained from: 
(i) an equation of state, $P=P(\rho)$; 
(ii) the equation of continuity

\begin{equation}
\partial_t\rho + \nabla\cdot\left[ \rho \bv \right] = 0 
\label{eq:Continuity}
\end{equation}
and 
(iii) from Newton's equation of motion,

\begin{equation}
\rho\left[ \partial_t \bv + \bv\cdot\nabla \bv \right] = -\nabla\cdot\sig^{(n)} - 
\gamma\nabla\cdot S + \bff \ .
\label{eq:Newton}
\end{equation}
Here $\bff$ is a local external force density acting on the system, e.g. due to stirring or interaction with the boundary. The flow equations (\ref{eq:Continuity}) and (\ref{eq:Newton}) evolve the density and velocity profiles of the system and they are valid wherever $\nabla \bv$ does not vanish. In ordinary liquids, it is often possible to replace the requirement of an equation of state by the simplifying assumption of incompressibility, but whether or not this is a good approximation for dense granular fluids is still an open question. 

\ni The flow equations are complete and self-consistent, but they are still short of a full description of the behaviour of granular fluids. The reason is that under dense flow conditions non-uniform velocity fields are unstable due to formation of regions of uniform flow, called plug flow \cite{Bi22, Jo70}, wherein $\bnabla\bv$ vanishes. Plug regions (PRs) are expected to form and grow in the absence of stirring forces and our next goal is to obtain the flow equations for the motion of PRs, as well as for their formation and expansion. 

\ni By definition, the acceleration field within a PR is spatially uniform; it is the total force on the region divided by its mass.
However, it should be noted that changes in the position of the boundaries of the PR must be accompanied by internal redistribution of stresses. The stress response rate does not enter the dynamics of the velocity field. We assume here that it is much faster than any other rate in the system and practically instantaneous.
The total force on a plug occupying a region $\Omega$, due to the normal forces applied on the region boundary $\partial\Omega$ by the adjacent fluid, is 

\begin{equation}
\bFF^n_P = - \int_{\partial\Omega} \sig^{(n)}_+(s)\cdot\bn(s) ds \ ,
\label{eq:NormalForce}
\end{equation}
where $ds$ is an infinitesimal surface element on $\partial\Omega$ and $\hat{n}(s)$  is an outward pointing unit vector normal to the boundary at point $s$. $\sig^{(n)}_+$ is the stress just at the outer side of the boundary $\partial\Omega$. The friction force acting on the region is given by

\begin{equation}
\bFF^f_P = - \gamma \int_{\partial\Omega} S_+(s)\cdot\bn(s) ds \ ,
\label{eq:FrictionForce}
\end{equation}
where $S_+$ is the value of $S$ also taken at the outer side of $\partial\Omega$. Due the distribution of intergranular contact orientations at the boundary of a volume element, the boundary force density may have normal and tangential components, both of which are captured by $S_+$. The total mass of the PR is

\begin{equation}
M_P = \int_\Omega \rho(\br) d^3\br 
\label{eq:PlugMass}
\end{equation}
and, since the term $\bv\cdot\bnabla \bv$ vanishes in $\Omega$, the PR acceleration is

\begin{equation}
\partial_t \bv_P = \left( \bF^n_P + \bF^f_P \right) / M_P \ . 
\label{eq:Accel}
\end{equation}
As we will demonstrate below, PRs are unstable and must grow. Therefore, to understand the flow behavior of dVF, we need to consider the kinematics of PR boundaries. To this end, it is convenient to define the scalar function 

\begin{equation}
\psi = \frac{1}{2} Tr\left( \tilT^2 \right) \ ,
\label{eq:Psi}
\end{equation}
which is finite only outside PRs. The growth of a PR is equivalent to an expansion of the external contour lines of zero $\psi$ surrounding it. The equation of motion for the scalar field $\psi$ is

\begin{equation}
\partial_t \psi + \bV_\psi\cdot\bnabla\psi = 0 \ .
\label{eq:PsiMotion}
\end{equation}
where $V_\psi(\br)$ is the velocity of the contour line of $\psi(\br)$ at location $\br$. Evidently, only the component of $V_\psi$ normal to $\partial\Omega$ is necessary to describe the expansion of the PR and this component is given by 

\begin{equation}
\bV^n_\psi = -\frac{\partial_t \psi\   \bnabla\psi}{\mid \bnabla\psi \mid^2} \ .
\label{eq:PsiNormal}
\end{equation}
The gradients on the right hand side of equation (\ref{eq:PsiNormal}) should be taken across the boundary and may be discontinuous. Nevertheless, a scrutiny of the numerator and the denominator will convince the reader that these discontinuities and the corresponding $\delta$-functions cancel out and leave a well-behaved term. This is also illustrated in the examples discussed in the following.

\ni To gain insight into the dynamics of growth of PRs in dVFs, we consider now an example that makes possible an explicit solution. Let a dVF be confined between two far-away boundaries at $x=\pm L$ and between $z=\pm\infty$ in the $z$-direction (figure 1a). We postulate an initial uniform density $\rho(x,t=0)=\rho_0$ and an arbitrary initial velocity profile in the $z$-direction, $v_0(x)\hat{z}=v(x,t=0)\hat{z}$. Compressive forces in the $x$-direction are applied uniformly across the boundaries at $x=\pm L$. These forces give rise to a stress whose only non-zero component is $\sig_{xx}$, taken to be constant. Due to the symmetry of the system and the boundary loading, neither the forces nor the initial conditions can change the density distribution, which alleviates the need to solve for the equation of state. 

\ni A straightforward example to analyse is when $v_0(x)$ is continuous and monotonic, i.e. $\partial_x v_0(x)\neq 0$ for all $x$. Then eq. (\ref{eq:Newton}) reduces to $\partial_t v = g$, whose simple solution is that the velocity profile remains constant with time, $v(x,t)=v(x,t=0)+gt$.

\ni A more interesting case is when the initial velocity profile has a local maximum, fixed at $x=0$ (figure 1b). 
A PR nucleates at the maximum, as described in detail in \cite{BlScEd08}. Basically, this is because the streamline at $x=0$ experiences friction forces opposing the flow from both sides and it must decelerate relative to its surrounding fluid. Eventually its velocity matches that of a neighbour streamline and they move together, forming a nucleus plug. The PR is slowed down by its surrounding fluid and continues growing by the same mechanism.
We wish to study the growth rate of the PR around the maximum. 
In regions not yet reached by the PR boundary, the velocity profile is monotonic and it does not change with time, as discussed above. Thus, for sufficiently short time $t$ the growth of the PR is dominated by the velocity profile near the maximum.
For convenience, we assume that the velocity can be expanded there as

\begin{equation}
v(x,t=0) = U - \alpha\left( x/x_0 \right)^2 + O\left[\left( x/x_0 \right)^3 \right] \ .
\label{eq:ParaProfileZero}
\end{equation}
The velocity profile at a later time $t$ is given by

\begin{equation}
v(x,t) = \left[v(x,t=0) + gt \right] \theta\left(t - t_p(x)\right) 
+ v_P(t) \theta\left(t_p(x) - t\right) \ ,
\label{eq:ParaProfileT}
\end{equation}
where $\theta$ is the Heavyside step function, $t_p$ is the time when the plug boundary reaches point $x$ and $v_p(t)$ is the velocity of the PR at time $t$. The location of the PR boundary at time $t$ is $l_p(t)$ (figure 1c) and the acceleration of the PR is 

\begin{equation}
\partial_t v_p = g - \frac{\gamma\sig_{xx}}{\rho_0 l_p(t)}  \ .
\label{eq:PlugAccel}
\end{equation}
The last term on the right of (\ref{eq:PlugAccel}) represents deceleration due to friction on the PR boundaries. 
The PR expands at a rate that can be found directly from (\ref{eq:PsiNormal}),

\begin{equation}
V_\Psi = \partial_t l_p = \frac{\gamma\sig_{xx}x^2_0}{2\rho_0 v_0 l^2_p}  \ .
\label{eq:PlugVelocity}
\end{equation}
Alternatively, (\ref{eq:PlugVelocity}) can be obtained by solving for the time $dt$ that it takes a streamline a distance $dx$ away from the boundary to match velocity with the PR. 
From (\ref{eq:PlugVelocity}) we obtain that the PR boundary grows as

\begin{figure}
\includegraphics[width=7cm]{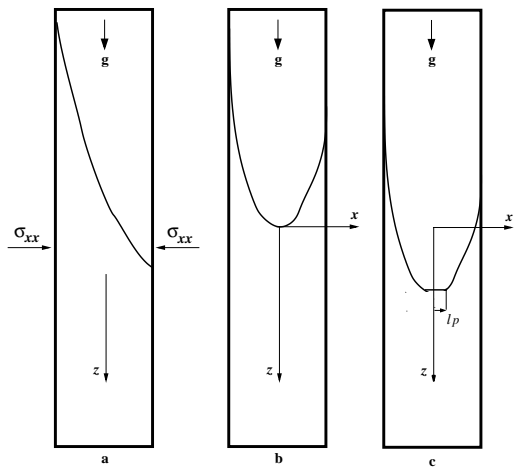}%
\caption{\label{Fig1} 
a) A simple uni-directional flow in the $z$-direction under gravity $g$. The initial downward velocity profile increases from left to right.  
b) Same as in $a$ but the velocity profile has a maximum. 
c) Nucleation and growth of a plug around a local maximum of the velocity profile. All the fluid accelerates at $g$ except for the stream line through the maximum, which is slowed down from both sides by friction forces. Consequently, a plug forms and grows around this point.}
\end{figure}


\begin{equation}
l_p(t) = \left[ 3\frac{\gamma\sig_{xx}x^2_0}{2\rho_0 v_0 }t \right]^{1/3} \ .
\label{eq:PlugPosition}
\end{equation}
Eqs. (\ref{eq:ParaProfileT}), (\ref{eq:PlugAccel}) and (\ref{eq:PlugPosition}) provide a full solution for the velocity profile. 
The expansion of PRs as $t^{1/3}$, described by (\ref{eq:PlugPosition}), is generic and can be understood as follows. 
The friction force is proportional to the surface area of the PR boundary, while the mass is proportional to its volume. 
In our example, the friction force that the PR experiences from its surrounding material is constant, while its mass increases proportionally to $l_p(t)$ as it expands. Consequently, the acceleration decreases inversely proportional to $l_p(t)$. 
Similarly, given an initial velocity profile $v_0(x,y,t=0)$ that is cylindrically symmetric around a local maximum, the friction per unit height experienced by a PR of height $h$ is proportional to $2\pi l_p(t)$, while its mass per unit height increases as $\pi l^2_p$. The reduction in acceleration is then also inversely proportional to $l_p$, leading again to the result that $l_p$ grows as $t^{1/3}$. It can be shown that the linear size of the region (however defined) increases as $t^{1/3}$ for any velocity profile that can be expanded around a local maximum at $(x_0,y_0)$ as 

\begin{equation}
v(x-x_0,y-y_0) = U - \alpha \left(\frac{(x-x_0)^2}{R^2_1} + \frac{(y-y_0)^2}{R^2_2}\right) + ...  \ .
\label{eq:VelocityProfGen}
\end{equation}

\ni To conclude, we have developed the flow equations of a fluid dominated by da Vinci - Amontons - Coulomb solid friction. 
The flow equations have been obtained for arbitrary flow regions, whether uniform or not. 
A key advantage of this model is that it gives rise naturally to unstable flows in the sense that the flow equations lead to formation and growth of plug regions. 
To our mind, this attribute of dense granular flow is extremely important but hitherto hardly studied. Our flow model is valid beyond these instabilities and it describes the formation, expansion and motion of the plugs region alongside the regions of nonuniform flow.
We have discussed simple cases that are amenable to analytic treatment and we have shown that a generic feature of the flow is that, once a plug region has formed, its linear size increases with time $t$ as $t^{1/3}$. 
It is interesting that this growth law of PRs, which we have found in a model for dense granular flows, has been observed also in simulations of muuch more dilute granular gases \cite{Third}. 

\ni The phenomena described here resemble strongly observations in flow of dense granular materials. Therefore, we propose this as a minimal model for such flows, when the material is dense and flows sufficiently slowly that grains maintain significant contact at all time. 
Experimental and numerical work, which we intend to take up in the future, is still needed to put this model to the test. In particular, it would be interesting to test the model's prediction for the growth rate of plug regions.


\bigskip
 
\begin{thebibliography}{99} 

\bibitem{Ra00} J. Rajchenbach, J. Granular Flows. Adv. Phys. {\bf 49}, 229 (2000). 
\bibitem{ZhKa96} T. Zhou and L.P. Kadanoff, Phys. Rev. {\bf E 54}, 623 (1996).
\bibitem{JaNaBe96} H.M. Jaeger, S.R. Nagel and R.P. Behringerâ Phys. Today {\bf 49}, 32 (1996).
\bibitem{Ha83} P.K. Haff, J. Fluid Mech. {\bf 134}, 401 (1983). 
\bibitem{LuSaJeCh84} C.K.K. Lun, S. B. Savage, D.J. Jeffrey and N. Chepurniy, J. Fluid Mech. {\bf 140}, 223 (1984). 
\bibitem{AzChMo99} E. Azanza, F. Chevoir and P. Moucheront, J. Fluid Mech. {\bf 400}, 199 (1999). 
\bibitem{Go99} I. Goldhirsch, Chaos {\bf 9}, 659 (1999).
\bibitem{Sa83} S.B. Savage, in {\it Advances in Micromechanics of Granular Materials}, eds. M. Satake and J.T. Jenkins (Elsevier, 1983). 
\bibitem{JoNoJa90} P.C. Johnson, P. Nott and R. Jackson, J. Fluid Mech. {\bf 210}, 501 (1990). 
\bibitem{NoJa92} P. Nott and R. Jackson, J. Fluid Mech. {\bf 241}, 125 (1992). 
\bibitem{AnJa92} K.G. Anderson and R. Jackson, J. Fluid Mech. 241, 145 (1992). 
\bibitem{Sa98} S.B. Savage, J. Fluid Mech. {\bf 377}, 1 (1998). 
\bibitem{LoBo0002} W. Losert, L. Bocquet, T.C. Lubensky, and J.P. Gollub, Phys. Rev. Lett. 85, 1428 (2000); 
L. Bocquet, W. Losert, D. Schalk, T.C. Lubensky, and J.P. Gollub, Phys. Rev. E 65, 011307 (2001).
\bibitem{Mi9900} P. Mills, D. Loggia and M. Tixier, Europhys. Lett. {\bf 45}, 733 (1999); P. Mills, M. Tixier and D. Loggia, Eur. Phys. J. {\bf E 1}, 5 (2000).
\bibitem{Bl04} R. Blumenfeld, Phys. Rev. Lett. {\bf 93}, 108301 (2004).
\bibitem{GeKrBl08} M. Gerritswen, G. Kreiss, R. Blumenfeld, Stress chain solutions in two-dimensional isostatic granular systems: fabric-dependent paths, leakage and branching, Phys. Rev. Lett., in print; M. Gerritswen, G. Kreiss, R. Blumenfeld, 
Analysis of stresses in two-dimensional isostatic granular systems Physica A, in print.
\bibitem{BoClLeOt01} J.-P. Bouchaud, P. Claudin, D. Levine and M. Otto, Eur. Phys. J. {\bf E 4}, 451 (2001).
\bibitem{Bi22} E.C. Bingham, {\it Fluidity and Plasticity} (McGraw-Hill, New York, 1922).
\bibitem{Jo70} A.M. Johnson, in {\it Physical Processes in Geology} (Freeman, Cooper and Co., San Francisco, 1970), pp. 433-534.
\bibitem{BoTa50} F.P. Bowden and D. Tabor, {\it The Friction and Lubrication of Solids} (Clarendon, Oxford, 1950).
\bibitem{dV} L. da Vinci, {\it Static measurements of sliding and rolling friction}, Codex Arundel, folios 40v, 41r, British Library.
\bibitem{Am1699} Amontons G., {\it Histoire de l'Academie Royale des Sciences avec les Memoires de Mathematique et de Physique, 1699-1708}, (Chez Gerald Kuyper, Amsterdam, 1706-1709), p. 206.
\bibitem{Co1779} C. A. de Coulomb, {\it Theorie des machines simples, en ayant egard au frottement de leurs parties et A la roideur des cordages}, (reprinted by Bachelier, Paris 1821).
\bibitem{BlScEd08} R. Blumenfeld, M. Schwartz and S. F. Edwards, The flow equations and catch-up dynamics of da Vinci Fluids, submitted.
\bibitem{Third} S. K. Das and S. Puri , Europhys. Lett. {\bf 61}, 749 (2003); S.K. Das and S. Puri, Phys. Rev. {\bf E 68}, 011302 (2003).

\end{thebibliography}
\end{document}